\documentclass[11pt]{article}
\usepackage[a4paper,margin=2.6cm]{geometry}
\usepackage[T1]{fontenc}
\usepackage{lmodern}
\usepackage{microtype}
\usepackage{amsmath,amssymb}
\usepackage{graphicx}
\usepackage{booktabs}
\usepackage[font=small,labelfont=bf]{caption}
\usepackage{xcolor}
\usepackage[colorlinks=true,linkcolor=blue!50!black,citecolor=blue!50!black,urlcolor=blue!50!black]{hyperref}
\newcommand{\Kq}{K_{\mathrm q}}

\title{How Many Keys Must a Quantum Computer Break\\ to Take Over a Blockchain?\\[0.4em]
\large A Quantum Nakamoto Coefficient for Fifteen Proof-of-Stake Networks}
\author{Seyyed Mohammad Hassan Halataei\\[0.3em]
\small Department of Physics, Shahid Beheshti University, Tehran, Iran\\
\small \href{mailto:m_halataei@sbu.ac.ir}{m\_halataei@sbu.ac.ir}}
\date{November 2026}

\begin{document}
\maketitle

\begin{abstract}
\noindent Blockchains that run on proof of stake, among them Ethereum and Solana, are secured by a vote: a transaction becomes irreversible once computers holding two thirds of the deposited money have signed it. Each signature is made with a secret key whose public counterpart is permanently visible, and a large quantum computer could compute the secret from the public key. We ask a simple question: how many such keys would it have to break to take over the vote? We call the answer the quantum Nakamoto coefficient and measure it on live data from fifteen networks, collected on 15 November 2026. For fourteen of them, 2 to 18 keys are enough to break the network's security guarantee and 7 to 77 to control it outright; at one key every 12 minutes, the pace implied by recent resource estimates, a single quantum computer would defeat the guarantee of any of these networks in less than four hours and take full control in less than sixteen hours. Ethereum is far more robust, but a 2025 change that lets operators merge their deposits into fewer keys has cut the number of keys at its first threshold from about 465,000 to 7,682, a factor of 60. We derive a simple formula for this effect, and another for the small rotating committee on which wallets rely, and confirm both on the data. For fourteen of the fifteen networks, the first large quantum computer would be less than a day away from total control. Their validator signatures should therefore be migrated to post-quantum schemes before such a machine exists, since once it does, the time to respond is measured in hours.
\end{abstract}


\section{Introduction}

A blockchain is a ledger kept in many copies with no central authority. In proof-of-stake networks, such as Ethereum and Solana, the copies are kept in agreement by a weighted vote. Participants called validators deposit money, their stake, and sign each new block of transactions with a digital signature; a signature counts in proportion to the stake behind it. Once validators holding more than two thirds of the stake have signed a block, it is final and cannot be reversed~\cite{casper,tm}.

A digital signature uses two keys: a private key that signs, and a public key, visible to everyone, with which a signature is checked. Its security rests on the difficulty of computing the private key from the public one. Shor's algorithm~\cite{shor} solves this problem efficiently on a quantum computer. Quantum hardware is still limited by noise and decoherence~\cite{leggett,thesis,hal17,hal25}, but rapid advances in quantum algorithms~\cite{pre,google} have reduced the requirement to fewer than 500,000 physical qubits on a superconducting machine and 18 to 23 minutes per key, or 9 to 12 minutes for a machine prepared in advance. Whoever holds a validator's private key can cast that validator's vote.

Most studies of this threat concern the theft of coins~\cite{agg,google}. The vote is the easier target: validator public keys are published and stay in use for months or years, so they can be attacked without any deadline. Reference~\cite{google} estimates that 20 machines would need more than nine months to obtain two thirds of the Ethereum vote, assuming about one million validators of equal weight, and notes that concentration of stake in large pools could shorten this; a recent survey~\cite{sok} discusses such attacks qualitatively. Neither counts how many keys are needed when stake is distributed over keys as it really is. That count is the subject of this paper. To our knowledge it has not been measured across networks, nor has the effect on it of Ethereum's recent merging of stake into fewer keys~\cite{eip}.

\section{Model}

Let a network have $N$ validator keys with voting weights $w_1\ge w_2\ge\dots\ge w_N$. Agreement is guaranteed only while less than one third of the weight is dishonest~\cite{casper,tm}. With more than $1/3$ an attacker can stop blocks from becoming final or, by signing two versions of a block, split the network between contradictory histories; with more than $1/2$ it decides which competing version prevails; and with more than $2/3$ it can make final whatever it chooses, with no help from anyone. We define the \emph{quantum Nakamoto coefficient} as the smallest number of keys whose weight exceeds a threshold $\theta$:
\begin{equation}
\Kq(\theta)=\min\left\{k\ :\ \sum_{i=1}^{k} w_i \;>\; \theta\sum_{i=1}^{N} w_i\right\}.
\label{eq:Kq}
\end{equation}

The name follows the Nakamoto coefficient~\cite{naka}, the number of organisations that would have to collude to cross the same threshold. The difference is the unit. An organisation that runs a thousand validators is one entity but a thousand separate problems for Shor's algorithm: a quantum attacker pays per key. Since each key costs the same time $\tau$, the best strategy is to break keys in order of decreasing weight. With $m$ machines the threshold is reached after
\begin{equation}
T_\theta=\frac{\Kq(\theta)\,\tau}{m}.
\label{eq:T}
\end{equation}
If the network replaced its keys at intervals $L$, the attack would succeed only with
\begin{equation}
m\;\ge\;\frac{\Kq(\theta)\,\tau}{L}
\label{eq:rot}
\end{equation}
machines. A stolen key does not silence its owner, who keeps voting honestly; the attacker adds contradictory votes in the owner's name.

\section{Data and results across networks}

On 15 November 2026 we downloaded from public servers of each network the list of active validators and their voting weights: eight networks built on the Cosmos software, Solana, NEAR, Aptos, Polygon, BNB Chain, TRON and Ethereum (2.41 million validator records), and the 19 signing keys of Wormhole, a bridge that moves assets between networks. In BNB Chain and TRON the 45 and 27 block producers have equal votes.

Table~\ref{tab:kq} lists $\Kq$ and the time $T$ for one machine at $\tau=12$ min, the value implied by the nine-month estimate of Ref.~\cite{google}. A full run of Shor's algorithm, the relevant figure when keys are broken one after another, takes 18 to 23 min, and on the 381-bit field of BLS12-381 perhaps twice as long~\cite{google}. All times scale as $\tau/m$; the key counts do not depend on $\tau$.

\begin{table}[ht]
\centering
\caption{Quantum Nakamoto coefficient at the three thresholds and time for one machine at 12 min per key, from data of 15 November 2026. $N$: number of signing keys. $^{a}$The same Ethereum stake held in 32-ETH keys, as before 2025. $^{b}$BLS12-381 for votes.}
\label{tab:kq}
\small\setlength{\tabcolsep}{3.6pt}
\begin{tabular}{llrrrrrr}
\toprule
Network & Key type & $N$ & $\Kq(1/3)$ & $\Kq(1/2)$ & $\Kq(2/3)$ & $T_{1/3}$ & $T_{2/3}$\\
\midrule
Kava & Ed25519 & 48 & 2 & 4 & 7 & 0.4 h & 1.4 h\\
Cosmos Hub & Ed25519 & 180 & 3 & 8 & 16 & 0.6 h & 3.2 h\\
Osmosis & Ed25519 & 30 & 4 & 7 & 12 & 0.8 h & 2.4 h\\
dYdX & Ed25519 & 21 & 4 & 7 & 10 & 0.8 h & 2.0 h\\
Injective & Ed25519 & 45 & 5 & 9 & 14 & 1.0 h & 2.8 h\\
Polygon PoS & secp256k1 & 100 & 4 & 7 & 11 & 0.8 h & 2.2 h\\
Celestia & Ed25519 & 98 & 6 & 13 & 25 & 1.2 h & 5.0 h\\
Axelar & Ed25519 & 41 & 8 & 14 & 20 & 1.6 h & 4.0 h\\
Sei & Ed25519 & 40 & 9 & 14 & 21 & 1.8 h & 4.2 h\\
NEAR & Ed25519 & 342 & 10 & 17 & 26 & 2.0 h & 5.2 h\\
TRON & secp256k1 & 27 & 10 & 14 & 19 & 2.0 h & 3.8 h\\
Aptos & BLS12-381 & 82 & 14 & 24 & 36 & 2.8 h & 7.2 h\\
BNB Chain & secp256k1, BLS$^{b}$ & 45 & 16 & 23 & 31 & 3.2 h & 6.2 h\\
Solana & Ed25519 & 662 & 18 & 39 & 77 & 3.6 h & 15.4 h\\
\midrule
Ethereum & BLS12-381 & 807{,}356 & 7{,}682 & 109{,}030 & 341{,}778 & 64 d & 7.8 yr\\
Ethereum, 32 ETH/key$^{a}$ & BLS12-381 & 1{,}396{,}487 & 465{,}496 & 698{,}244 & 930{,}992 & 10.6 yr & 21.2 yr\\
\bottomrule
\end{tabular}
\end{table}

\begin{figure}[ht]
\centering
\includegraphics[width=0.85\textwidth]{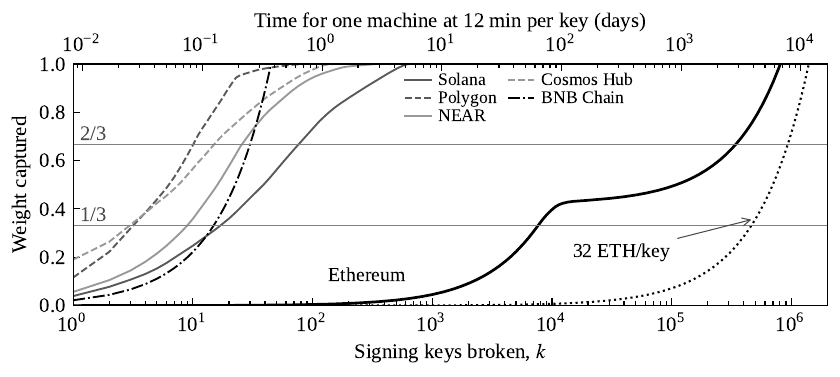}
\caption{Share of the vote captured versus number of keys broken, largest first. Upper axis: time for one machine at 12 min per key. Dotted: Ethereum with the same stake held in 32-ETH keys.}
\label{fig:capture}
\end{figure}

Outside Ethereum, $\Kq(1/3)$ lies between 2 and 18 and $\Kq(2/3)$ between 7 and 77. The number of validators is a poor guide: the Cosmos Hub has 180, yet three keys hold more than a third of the stake, the largest alone 19\%. One machine would therefore cross the first threshold of any of these networks in less than four hours and the last in less than sixteen hours (Fig.~\ref{fig:capture}). The Wormhole bridge accepts a transfer signed by 13 of its 19 keys: 2.6 hours. Because the keys do not change, a slower machine alters the calendar but not the conclusion: at one key per day, every network in Table~\ref{tab:kq} except Ethereum reaches the $1/3$ threshold within 18 days.

\section{Ethereum: merged stake and the wallet committee}

\subsection{Merged stake}
Ethereum needs hundreds to thousands of times more keys, but its margin is shrinking. Until 2025 a validator key could carry only 32 ETH, so a large operator needed thousands of keys. The protocol now allows up to $B=2048$ ETH per key~\cite{eip}, and operators are merging their validators to reduce network load. In our data 807,356 keys hold 44.69 million ETH; the 9,525 keys with at least 1024 ETH hold 38.8\% of it, and 778,891 keys of 32 ETH hold 55.8\%.

Suppose a fraction $c$ of the total stake $S$ sits in merged keys of size $B$ and the rest in keys of size $b=32$ ETH. Counting the large keys first, Eq.~\eqref{eq:Kq} gives
\begin{equation}
\Kq(\theta)=
\begin{cases}
\dfrac{\theta S}{B}, & \theta\le c,\\[10pt]
\dfrac{cS}{B}+\dfrac{(\theta-c)\,S}{b}, & \theta>c.
\end{cases}
\label{eq:merge}
\end{equation}
With $c=0.388$, Eq.~\eqref{eq:merge} predicts $\Kq(1/3)=7{,}273$. The data give 7,682, against 465,496 if the same stake were held in 32-ETH keys. The $1/3$ threshold now lies entirely within the merged stake, and its cost has fallen by a factor of 60, close to the limit $B/b=64$. One machine needs 64 days, 20 machines 3.2 days. The $2/3$ threshold still requires small keys: Eq.~\eqref{eq:merge} gives 397,800 and the data 341,778 (keys of intermediate size explain the difference), or 142 days for 20 machines instead of about nine months for equal validators. Every merger raises $c$; once $c$ exceeds $2/3$, the $2/3$ threshold also falls 64-fold, to about 14,500 keys at the present stake.

\subsection{The wallet committee}
Phones, wallets and bridges cannot follow 800,000 validators. They rely on a jury, the sync committee: $n=512$ validators drawn at random every 27.3 hours, with probability proportional to stake, and they accept what $n_{\mathrm q}=342$ of the jurors sign~\cite{altair}. If the attacker holds the keys of a fraction $p$ of the stake, the number of jurors it controls follows a binomial distribution, and it controls the jury of a given period with probability
\begin{equation}
P_{\mathrm{cap}}(p)=\sum_{j=n_{\mathrm q}}^{n}\binom{n}{j}\,p^{j}\,(1-p)^{\,n-j}.
\label{eq:cap}
\end{equation}

\begin{figure}[ht]
\centering
\includegraphics[width=0.75\textwidth]{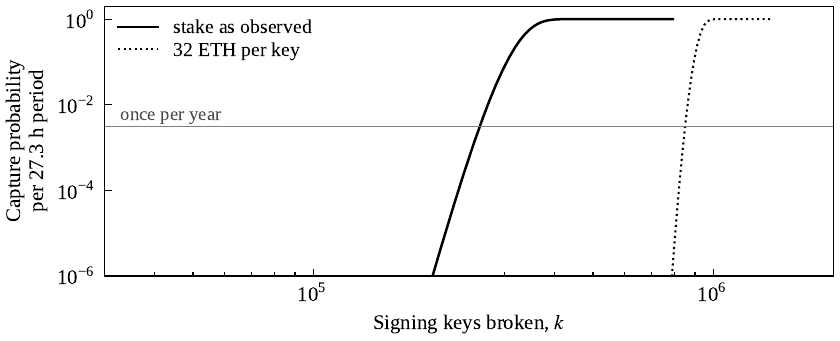}
\caption{Probability, Eq.~\eqref{eq:cap}, that the attacker controls at least 342 of the 512 committee seats in one 27.3-hour period, versus the number of keys broken. Solid: stake as observed; dotted: 32 ETH per key.}
\label{fig:jury}
\end{figure}

The sampling rule can be tested directly: keys of at least 1024 ETH should fill $512\times0.388=198\pm11$ seats, and the committee in office when we took the data contained 193. Equation~\eqref{eq:cap} rises steeply (Fig.~\ref{fig:jury}): $P_{\mathrm{cap}}$ is $10^{-14}$ at $p=0.50$, $9\times10^{-4}$ at $0.60$ and $0.50$ at $0.667$. The attacker expects to control one jury per year once $p=0.61$, which takes about 260,600 keys. The jury thus gives wallets nearly the two-thirds margin of the full vote. It is weaker in two respects: the keys of a past jury remain useful against a device that has been offline since, and jurors who sign falsely are not penalised.

\section{Discussion}

Three points deserve comment. First, in most proof-of-stake networks the vote, not the coins, is the cheapest target for a quantum computer: a few tens of keys, all public, with no deadline.

Second, changing keys regularly helps only if the interval $L$ is shorter than $\Kq\tau/m$, Eq.~\eqref{eq:rot}. For the small networks of Table~\ref{tab:kq} this means hours; for Ethereum's first threshold it means weeks against one machine, which the fast rotation proposed in Ref.~\cite{google} could achieve, but days against twenty. Short-lived keys tied to a quantum-safe identity~\cite{tqr} help, but the lasting remedy is post-quantum signatures~\cite{fips,xmss}.

Third, decentralisation counted in organisations and counted in keys can move in opposite directions: merging stake leaves the set of operators unchanged but divides the attacker's work by up to 64. For Ethereum this is a reason to complete the change of signatures before the merged share $c$ approaches $2/3$, starting with the largest keys.

The limitations are these. The weights are those of a single day, and about 2\% of the Ethereum keys counted belong to validators entering or leaving. The time per key is taken from Ref.~\cite{google}, not computed here. We count keys as published; an operator that keeps many keys in one place may be vulnerable by other means. We consider only Shor's algorithm on an error-corrected computer; Ising machines and quantum annealers, built for optimisation problems, are not known to threaten elliptic-curve keys. Proof-of-work networks such as Bitcoin have no voting keys, and protocols with sampled voting, such as Avalanche, fall outside Eq.~\eqref{eq:Kq}.

\section{Conclusion}

We asked a simple question: how many signing keys must a quantum computer break to take over the vote that secures a proof-of-stake blockchain? We called the answer the quantum Nakamoto coefficient and measured it on public data from fifteen networks.

The answer is unexpectedly small. In fourteen networks, 2 to 18 keys are enough to break the guarantee that transactions are final, and 7 to 77 are enough for full control. For a machine that needs 12 to 23 minutes per key this is a matter of hours; a machine a hundred times slower would still cross the first threshold in days or weeks, because the keys are public, long-lived and can be attacked with no deadline. What counts is not the number of validators but how the stake is spread over keys.

Ethereum is the exception, with hundreds to thousands of times more keys to break. But the merging of stake into large keys, allowed since 2025, has already lowered its first threshold sixty-fold, to 7,682 keys, about two months of work for one machine. A simple formula, Eq.~\eqref{eq:merge}, reproduces this number and predicts a further fall as merging continues. The 512-member committee on which phones and wallets rely obeys a binomial law, Eq.~\eqref{eq:cap}, confirmed on the committee in office.

Three practical implications follow. Networks with a few heavy keys should treat the replacement of their voting signatures by quantum-safe ones as at least as urgent as the protection of coins. Ethereum should complete that replacement before more stake is merged, beginning with the largest keys. And the coefficient, which anyone can recompute from public data in minutes, gives users, investors and regulators one number for comparing networks.

No quantum computer capable of this exists today, and the time per key is a published estimate. But the margin the data reveal is a few dozen keys and less than a day, and it will not widen by itself. Validator signatures should therefore be migrated to post-quantum schemes before such a machine exists, since once it does, the time to respond is measured in hours.

\paragraph{Data and code.} All numbers and figures can be reproduced from public network data with scripts available from the author.

\end{document}